\documentclass[reprint,amsmath,amssymb,prl,twocolumn,nofootinbib]{revtex4}
\usepackage{amsfonts}    
\usepackage{amssymb}
\usepackage{latexsym}
\usepackage{graphicx}
\usepackage{rotating}
\usepackage{multirow}
\usepackage[usenames,dvipsnames]{color}
\usepackage[active]{srcltx}

\usepackage{hyperref}
\usepackage{color}

\newcommand{\rar}{\rightarrow}

\newcommand{\non}{\nonumber}

\begin{document}


\title{Particular superintegrability of 3-body (modified) Newtonian Gravity}
\date{\today}

\author{Alexander~V.~Turbiner}
\email{turbiner@nucleares.unam.mx}
\author{Juan Carlos Lopez Vieyra}

\email{vieyra@nucleares.unam.mx}
\affiliation{Instituto de Ciencias Nucleares, Universidad Nacional
Aut\'onoma de M\'exico, Apartado Postal 70-543, 04510 M\'exico,
D.F., Mexico{}}

\begin{abstract}

It is found explicitly 5 Liouville integrals in addition to total angular momentum
which Poisson commute with Hamiltonian of  3-body  Newtonian Gravity in
${\mathbb R}^3$ along the Remarkable Figure-8-shape trajectory discovered by Moore-Chenciner-Montgomery.
It is checked that they become constants of motion along this trajectory. 
Hence, 3-body choreographic motion on Figure-8-shape
trajectory in ${\mathbb R}^3$ Newtonian gravity (Moore, 1993), as well as in  ${\mathbb R}^2$
modified Newtonian gravity by Fujiwara et al, 2003, is maximally superintegrable. It is conjectured that any  3-body potential theory which admit Figure-8-shape choreographic motion is superintegrable along the trajectory.

\end{abstract}

\maketitle

In 1993 C. Moore \cite{Moore:1993} discovered a new phenomenon in  3-body classical
chaotic dynamics: three particles of equal masses in ${\mathbb R}^3$
Newtonian gravity can perform periodic motion on the same Figure-8-shape closed
trajectory with zero total angular momentum and equal time delay without collisions. Later
it was confirmed rigorously by Chenciner-Montgomery \cite{CM:2000}, who also proved
that the trajectory is non-algebraic curve; it was also demonstrated its certain stability
towards small perturbations of initial data \cite{CGMS}.
This phenomenon was called {\it choreography} \cite{Simo}. It was studied intensely since
2000 for different number of particles and various pairwise potentials being mostly limited
to planar trajectories. It manifests the existence of {\it moving} equilibrium configuration: at
least, for attractive pairwise potentials the motion can not be stopped due to a possibility of
collapse otherwise, thus, a standard steady equilibrium configuration at zero velocities
does not exist.
The goal of the present Letter is to demonstrate that  3-body choreography on the
Remarkable Figure Eight trajectory, as was called in \cite{CM:2000}, in
${\mathbb R}^3$ Newtonian Gravity is characterized by 7 constants of motion, which is the
maximal possible amount, hence, being maximally superintegrable. Furthermore, there exist
6 independent functions on phase space which Poisson-commute with Hamiltonian along
this trajectory. Hence, the Hamiltonian is maximally superintegrable along the Remarkable
Figure Eight trajectory. As the first step we consider a ${\mathbb R}^2$
modified Newtonian gravity theory proposed by Fujiwara et al, \cite{Fujiwara:2003}, which allows
the  3-body choreographic motion on {\it algebraic} lemniscate by Jacob Bernoulli (1694),
and we construct explicitly seven, polynomial in coordinates and momenta, constants of
motion along the trajectory.

We begin by introducing the notion of {\it particular} integral \cite{Turbiner:2013} in
classical mechanics. Take a function $I(p,x)$ on classical phase space $(p,x)$, define the
Hamiltonian $H(p,x)$ and calculate the Poisson bracket $\{H, I\}$. If for some
trajectory(ies) in phase space the Poisson bracket vanishes, the function $I(p,x)$ is called a
particular Liouville integral, it becomes constant (of motion) on the trajectory: the
corresponding trajectory is called {\it particular}.
Otherwise, if the Poisson bracket vanishes for any trajectory, $I(p,x)$ is a Liouville or global
integral; it implies the r.h.s. of the Poisson bracket is zero identically. In order to have a
non-trivial dynamics the maximal number of functionally independent integrals (including the
Hamiltonian) should not exceed the dimension of phase space $d_{ph}=2d_c$ minus one.
When the number of integrals is bigger than the dimension of the coordinate space $d_c$, the
dynamics is called superintegrable. It is known that in the case of maximal superintegrability
where all $(d_{ph}-1)$ integrals are global, all bounded trajectories in coordinate space are
closed (and periodic) \cite{Nekhoroshev:1972}, as for a concrete example see for instance
\cite{TTW:2010}.
It is natural to attempt to generalize the statement of \cite{Nekhoroshev:1972}: even though
some integrals are particular and emerging constants of motion occur for the same particular
trajectory, the maximal {\it particular} superintegrability leads to closed particular trajectory.
It will be shown that for  3-body Newtonian choreography and one on algebraic lemniscate
it is the case. Note that following the well-known H.~Poincare theorem the  3-body
Newtonian dynamics in relative space is characterized by two global integrals, energy and
total angular momentum, only.

\noindent
{\it (A)\ Modified ${\mathbb R}^2$ Newtonian gravity: the example of
inverse problem.}  In 2003 Fujiwara et al, \cite{Fujiwara:2003} solved the inverse problem:
by taking the algebraic lemniscate on $(x,y)$-plane
\begin{equation}
\label{lemn}
 (x^2+y^2)^2\ =\ c^2 (x^2-y^2)\ ,
\end{equation}
as the trajectory, where without loss of generality one can take $c=1$, it was shown that is a
potential problem for 3 unit mass, point-like particles subject to pairwise potentials,
\begin{equation}
\label{potential}
 V\ =\ \sum_{i<j}^3 \left\{  \frac{1}{4}{\ln r_{ij}^2}  - \frac{\sqrt{3}}{24}  r_{ij}
^2\right\}
 \ \equiv \ \frac{1}{4}\, \ln I_1 - \frac{\sqrt{3}}{24}\, I_2\ ,
\end{equation}
at zero angular momentum. Here
\[
 r_{ij}=\sqrt{( {\mathbf x}_i -{\mathbf x}_j )^2}\ ,
\]
is relative distance between particles $i$ and $j$.

\begin{figure}[!htb]
  \begin{center}
\includegraphics[width=3.0in]{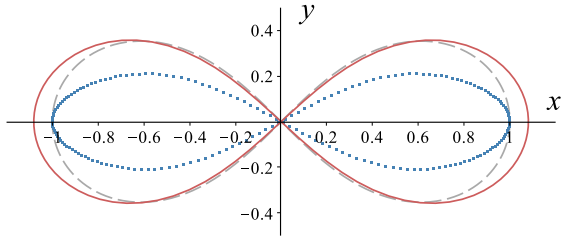}
    \caption{
    \label{Figure}
    Figure Eight trajectories for 3 unit mass particles: algebraic lemniscate
    - grey (dashed) line, ${\mathbb R}^2$ gravity -blue (dotted) line,
    ${\mathbb R}^3$ gravity - red (solid) line
    }
  \end{center}
\end{figure}

The first attractive terms in (\ref{potential}) represents  3-body ${\mathbb R}^2$
Newtonian potential at gravitational constant $G=1/2$, while the second repulsive term is
square of hyperradius in the space of relative motion. Each pairwise potential in
(\ref{potential}) has maximum at $r^{(max)}_{ij}=1.8612$. It can be shown that the
motion on the trajectory is confined to the domain
$r_{ij}^{(min)}\leq r_{ij} \leq r^{(max)}_{ij}$, where $r_{ij}^{(min)}=0.6813$,
thus, it never enters into the domain of configuration space, where the repulsion
term $\sim I_2$ dominates.

It is known that the algebraic lemniscate (\ref{lemn}) is parametrized by
\begin{equation}
\label{param}
 x(t)\ =\ c\, \frac{{\rm sn}(t,k)}{1+{\rm cn}^2(t,k)}\quad ,\quad
 y(t)\ =\ c\, \frac{{\rm sn}(t,k)\, {\rm cn}(t,k)}{1+{\rm cn}^2(t,k)}\ ,
\end{equation}
see e.g. \cite{Fujiwara:2003}, where ${\rm sn}(t,k),\ {\rm cn}(t,k)$ are Jacobi elliptic
functions,
$k \in [0,1]$ is elliptic modulus, which defines via complete elliptic integral, the real period
$T = 4 \int_0^1 \frac{dx}{\sqrt{(1-x^2)(1-k^2x^2)}}$: $x(t)=x(t+T), y(t)=y(t+T)$. The
evolution is defined by
\begin{align}
\label{evolution}
  x_1(t) = x(t) \quad \,, \quad
  y_1(t) &= y(t) \ ,  \non
\\
  x_2(t)  = x(t + {T}/{3})\quad \,, \quad
  y_2(t) &= y(t + {T}/{3}) \ ,
\\
  x_3(t) =  x(t - {T}/{3}) \quad \,, \quad
  y_3(t) &= y(t - {T}/{3}) \ , \non
\end{align}
for the first, second and third particles, respectively. By straightforward calculation using
Maple 18 in 20-digit arithmetic one can check that the center-of-mass is conserved
\[
  {\mathbf X}_{\rm CM}(t) = {\mathbf x}_1 + {\mathbf x}_2 + {\mathbf x}_3 =0\ ,
\]
and fixed
\[
  {\mathbf V}_{\rm CM}(t) = {\mathbf v}_1 + {\mathbf v}_2 + {\mathbf v}_3 =0\ ,
\]
for elliptic modulus
\[
 k_0^2= \frac{2+\sqrt{3}}{4} = \left(\frac{1+\sqrt{3}}{2\sqrt{2}}\right)^2 \
 \simeq\ {0.933012701892}  \ldots \ ,
\]
only \cite{Fujiwara:2003}. It corresponds to period $T=11.07225\ldots$
There exist times along the evolution, given by  \hbox{$t=\frac{T}{2}n\,, $}
\hbox{$ n=0,1,2,\ldots$}, when all three particles
are on a straight line - it is called the {\it Euler} line \cite{Simo} - which forms the {\it Euler}
angle with horizontal $x$-coordinate (the line of symmetry $y \rar -y$) equal to $15^o$, see
Fig.1, where usually initial conditions are defined.  Note that for some instance
between two consecutive Euler linear configurations the positions of the bodies correspond to the vertices
of an isosceles triangle with one of the bodies located on the $x$-axis. Sometimes this configuration
is used to introduce initial data.  Let us consider, the following ten, polynomial in coordinates and velocities,
expressions  numbered in italics,
\begin{enumerate}
\item[(\it 1)]\ $L=\sum {\mathbf x}_i \times {\mathbf v}_i = 0$  Angular Momentum\,,
\item[(\it 2)]\ $E={T} + {V}$   Total Energy  (dependable) \\
\( \displaystyle =\frac{1}{4}\log\left(\frac{3\sqrt{3}}{2}\right) \simeq 0.23869   \)\,,
\item[(\it 3)]\ $\displaystyle I_1 = r_{12}^2 \,r_{13}^2 \,r_{23}^2 =   \frac{3\sqrt{3}}{2}$
\,,
\item[(\it 4)]\ $\displaystyle I_2 = 3\sum {\mathbf x}_i^2 = \sum_{i<j} r_{ij}^2 =
3\sqrt{3}\qquad $\\  Moment of Inertia\,,
\item[(\it 5)]\ $\displaystyle {T}= \frac{1}{2}\sum {\mathbf v}_i^2 =   \frac{3}{8}\qquad$
Kinetic Energy\,,
\item[(\it 6)]\  $\displaystyle {\tilde {T}} = {\mathbf v}_1^2 \, {\mathbf v}_2^2 \,{\mathbf
v}_3^2  =   \frac{1}{128}$\,,
\item[(\it 7-10)]\ $J_i(k) = {\mathbf v}_i^2 + (k^2 -\frac{1}{2}) {\mathbf x}_i^2 = \frac{1}
{2}\ , \ i=1,2,3$\ ,\\
$\displaystyle J_s \equiv \sum_{i=1}^3 J_i(k_0) = 2\ {T} + \frac{1}{3}\left(k_0^2- \frac{1}
{2}\right)I_2 = \frac{3}{2}$\ .
\end{enumerate}
 It can be shown that all ten expressions from {\it (1)-(10)} become constants
under the evolution (\ref{param}), (\ref{evolution}).
Seven of them are functionally (algebraically) independent.
Let us choose $L, I_1, I_2, {T}, \tilde{T}, J_1, J_2$\,.
It can be shown that in spite of polynomiality of these expressions
they do not generate infinite-dimensional, finitely-generated polynomial
Poisson algebra with 7 generators as "seeds" as it might be expected. However, 5 of them
$L, I_1, I_2, {T}, \tilde{T}$ do generate a commutative algebra in the following sense: all
Poisson brackets being calculated on the algebraic lemniscate (\ref{lemn}) vanish.

The Hamiltonian
\begin{equation}
\label{HF}
{\cal H}={T} + \frac{1}{4}\, \ln I_1 - \frac{\sqrt{3}}{24}\, I_2\ ,
\end{equation}
see (\ref{potential}), is made from particular integrals $T, I_1, I_2$. It can be shown
explicitly that the Poisson bracket $\{{\cal H}, L\}=0$ and r.h.s. of the five Poisson brackets
vanish on algebraic lemniscate (\ref{lemn}):
\[
 \{{\cal H}, T\}=\{{\cal H}, I_1\}=\{{\cal H}, I_2\}=\{{\cal H}, \tilde T\}=\{{\cal H}, J_s\}
=0\ ,
\]
noting that $J_s$ is made from $T, I_2$, see {(\it 10)}.
Hence, the Hamiltonian ${\cal H}$ is particularly superintegrable, where $L$ is global and
$T, I_1, I_2, \tilde T$ are four particular Liouville integrals. We were unable to find
particular Liouville integrals, for which $J_{1,2,3}$ become eventually constants of motion
on trajectory (\ref{lemn}).
Interestingly, it can be shown that for any positive $c$ in (\ref{param}) the energy $E$ is
positive and when $c$ tends to zero the energy vanishes.

We conclude that the { 3-body} choreographic motion with pairwise potential
(\ref{potential}) is maximally (particularly) superintegrable: there exist 7 constants of
motion. Taking velocity-independent integrals (constants of motion) {(\it 3)} and  {(\it 4)}
one can see that the { 3-body} choreographic motion in the ${\mathbb R}^3$ configuration
space parametrized by squared relative distances follows a planar closed curve. The same is
true in momentum space when coordinate-independent integrals (constants of motion) {(\it
5)} and  {(\it 6)} are taken. Both curves are elliptic ones.
Thus, the closed periodic trajectory in phase space is factorized: two closed periodic
(elliptic)  curves, one is in coordinate space and another one is in momentum space, factor
out.

It has to be noted that 3-body choreographic figure-8-shape motion exists for
${\mathbb R}^2$ Newtonian gravity, see Fig.1, at gravitational constant
$G=1/2$, when the second term in Fujiwara et al, potential (\ref{potential}) vanishes, $I_2=0$. 
The figure-8-shape curve is known numerically, see \cite{Fukuda}. 
This choreography is characterized by the period $T=6.904470$. The Euler line also exists in this case, its Euler angle (with $x$-coordinate) is $\sim 8.90267^{o}$. In general, the question about its (particular) superintegrability remains open, although total angular momentum is conserved as well as modified ${\cal I}_1$ (see below, eq.(8)).

\noindent
{\it (B)\  3-body  ${\mathbb R}^3$ Newton problem.} The
Hamiltonian for three point particles of unit mass is given by,
\begin{equation}
\label{Hamiltonian-Moore}
  {\cal H}\ =\ \frac{1}{2}\sum^3_{i=1} {\mathbf v}_i^2 - \sum^3_{i < j}
\frac{1}{r_{ij}}\ ,
\end{equation}
where the gravitational constant is assumed equal to one, $G=1$. Under special initial data
the system of four coupled Newton equations in space of relative motion leads to
choreographic figure-8-shape trajectory as a solution when three particles move one after
another with equal time delay, see Fig.1. The Euler line exists, its Euler angle is $\sim
14.0688^{o}$, for details see \cite{Simo}. This trajectory (and its initial data) is known
numerically with high accuracy \cite{Simo}.
3-body evolution can be constructed approximately, in particular, by using
Mathematica code designed by Moeckel \cite{Moeckel},
unlike the case of algebraic lemniscate, where it is known in terms of Jacobi elliptic function, see (\ref{evolution}).
Using the code \cite{Moeckel}, it is found the period $T=6.325913\ldots$ and checked that
both global integrals, energy and total angular momentum, are constants of motion,
\begin{equation}
\label{3-body: EL}
  E\ = \ -1.287142\ ,\ L\ =\ 0.000000\, ,
\end{equation}
with accuracy of 7 figures.  The next decimal digit in (\ref{3-body: EL}) floats
being time dependable.
The important guess, based on Weierstrass polynomial approximation theorem
about approximation a function by polynomials, is to look for possible particular Liouville
integral for (\ref{Hamiltonian-Moore}) in a form of single argument Taylor expansion
taking
as entry the particular integrals of motion for algebraic lemniscate (\ref{lemn}) in the
potential (\ref{potential}): $I_1, I_2, T, {\tilde T}, J_s$, see below ({\it 1})-({\it 10}).
Following this guess we manage to construct all 5 particular Liouville integrals, which
Poisson-commute with the Hamiltonian (\ref{Hamiltonian-Moore}): ${\cal I}_1, {\cal I}_2,
{\cal T}, \tilde{\cal T}, {\cal J}_s$, see (\ref{I1N}), (\ref{I2N}), (\ref{T3N}), (\ref{T4N}),
(\ref{J5N}), on Remarkable figure-8-shape trajectory. Right-hand-side of Poisson brackets is
 almost zero being of order of $10^{-7}$ (or less) in agreement with accuracy
provided by the evolution \cite{Moeckel},  see (\ref{3-body: EL}).
In order to check consistency it is verified these particular Liouville integrals on
Remarkable figure-8-shape trajectory become constants,
\begin{widetext}
\begin{align}
 {\cal I}_1\ =&\ I_1 \ -\ 0.5450972\, I_1^2\ +\ 0.13047597\, I_1^3\ -
      \ 0.00514404\, I_1^4\ -\ 0.003987607 \, I_1^5 \ +               \label{I1N} \\
   &  \ 8.130425 \times 10^{-4}\, I_1^6 \ -\ 5.441899 \times 10^{-5}\, I_1^7\ +
      \ 6.4018378 \times 10^{-7}\, I_1^8 \  \thickapprox \ 0.7092995\ ,
\nonumber\\[5pt]
{\cal I}_2\ =&\ I_2 - 0.2241790\, I_2^2\ +\ 0.0150429\, I_2^3\ +
     \ 0.000838\, I_2^4 \ -\ 0.0001117\, I_2^5\ +  \label{I2N} \\
   & \ 0.00002001\,I_1^6\ - 4.821201\times 10^{-6}\, I_2^7\ +
     \ 3.15375\times 10^{-7}\, I_2^8  \ \thickapprox \ 1.5099589\ ,
\nonumber \\[5pt]
 {\cal T}_3\ =&\ {T}\ -\ 1.7073653\, {T}^2\ +\ 1.27576\, {T}^3\ -\
       0.0937725 \, {T}^4\ -\ 0.53251342 \, { T}^5\ +  \label{T3N} \\
   & \ 0.38893304 \, {T}^6\ -\ 0.1161555\, {T}^7 +
     \ 0.0131440 \, {T}^8 \  \thickapprox\ 0.2280553\ ,
\nonumber\\[5pt]
 {\tilde {\cal T}}_3\ =&\ {\tilde {T}}\ -\ 11.627099 \, {\tilde { T}}^2\ +
     \ 77.1466332\, {\tilde { T}}^3\ -\ 319.485185 \, {\tilde { T}}^4\ +\ 845.607914 \, {\tilde
{ T}}^5\ -  \label{T4N} \\
   &   1396.91200 \, {\tilde { T}}^6\ +\ 1316.83205 \, {\tilde { T}}^7\ -\
       542.334849 \, {\tilde { T}}^8\  \thickapprox\ 0.03757715\ ,
\nonumber\\[5pt]
 {\cal J}_s\ =&\ J_s\ -\ 1.12957370\, J_s^2\ +\ 0.72387170\, J_s^3\ -\ 0.28746375\, J_s^4\ +\
       7.2313705 \times 10^{-2}\, J_s^5\ -                                                        \label{J5N} \\
   &   1.12265877\times 10^{-2}\, J_s^6\ +\ 9.801857\times 10^{-4}\, J_s^7\
   -\  3.66702 \times 10^{-5} \, J_s^8 \ \thickapprox\ 0.38492968\ . \nonumber
\end{align}
\end{widetext}
Amplitudes of oscillations of the constants of motion are of the order of $10^{-7}$ or less, it
implies that the last figure in (8)-(12) can be changed in one unit. Thus, the 3-body
choreography with zero total angular momentum in ${\bf R}^3$ Newton gravity is
maximally, particularly, superintegrable being characterized by 7 constants of motion $E, L,
{\cal I}_1, {\cal I}_2, {\cal T}, \tilde{\cal T}, {\cal J}_s$. Hence, the 3-body ${\bf R}^3$
Newton gravity is maximally particularly superintegrable being characterized by 6 (one
global and five particular) Liouville integrals.

\bigskip

\noindent
{\it Conclusions.} 3-body choreography with zero angular momentum exists for a number of
pairwise potentials both attractive at all distances and also repulsive at small distances being
attractive at large distances only \cite{Fukuda}. We conjecture that all of them are
maximally particularly superintegrable: it might be the intrinsic property of choreography
explaining its existence.

It is already known that 5-body choreography on algebraic lemniscate, found in
\cite{Fujiwara:2004}, is potential problem for two values of elliptic moduli in (\ref{param})
with pairwise potential \cite{JC:2019}. It is characterized by 15 explicitly-found Liouville
integrals which become the constants of motion on algebraic lemniscate, hence, the
trajectory is maximally, particularly superintegrable \cite{TL:2019}. 5-body choreography
on Remarkable figure-8-shape trajectory in ${\mathbb R}^3$ Newton gravity also exists
\cite{Simo}. The question about its integrability is open.

The phenomenon of choreography manifests the appearance of a new type of equilibrium
configurations: moving, non-steady equilibrium.\\

\noindent
{\it Acknowledgements.} The authors thank T Fujiwara, R Moeckel, R Montgomery and C
Sim\'o for useful mail correspondence and in some cases for personal discussions (R.M. and
R.M.) and
T Damour (IHES) for the important remark.
A.V.T. is grateful to participants of the seminars at University of Minnesota, Simons Center
for Geometry and Physics, C.N. Yang Institute for Theoretical Physics and Stony Brook
University, all at Stony Brook, especially, to V Korepin and R Schrock for interest to this
work.
This research is partially supported by CONACyT A1-S-17364 and DGAPA IN113819
grants (Mexico).


\end{document}